\documentclass[10pt, twocolumn, aps, prl, superscriptaddress, nofootinbib]{revtex4-2}
\usepackage{bm}
\usepackage{amsmath, amssymb, braket}
\usepackage{graphicx}
\usepackage{upgreek}
\usepackage{soul}
\usepackage[normalem]{ulem}
\usepackage[unicode=true, pdfusetitle, bookmarks=true,
            bookmarksnumbered=false, bookmarksopen=true,
            bookmarksopenlevel=1, breaklinks=false,
            pdfborder={0 0 1}, backref=false,
            colorlinks=true]{hyperref}
\hypersetup{pdfborderstyle=, linkcolor=blue, 
            urlcolor=blue, citecolor=blue}
            
\hypersetup{
    pdftitle={Cavity-Mediated Collective Resonant Suppression of Local Molecular Vibrations},
    pdfauthor={Vasil Rokaj, Ilia Tutunnukov, H. R. Sadeghpour},
    pdfsubject={Research article},
    pdfkeywords={Hopfield, polaritons, ultrastrong coupling, 
                 vibrational strong coupling, VSC},
    pdfstartview={FitH},
    pdfpagemode={UseOutlines},
    pdfencoding={auto},
    breaklinks=true
}

\usepackage{orcidlink}

\newcommand{\rmv}{\mathrm{v}}

\begin{document}

\title{Cavity-Mediated Collective Resonant Suppression of Local Molecular Vibrations}

\author{Vasil Rokaj\,\orcidlink{0000-0002-0627-7292}}
\email{vasil.rokaj@villanova.edu}
\affiliation{Department of Physics, Villanova University, 
             Villanova, Pennsylvania 19085, USA \looseness=-1}

\author{Ilia Tutunnikov\,\orcidlink{0000-0002-8291-7335}}
\affiliation{ITAMP, Center for Astrophysics $|$ Harvard \& Smithsonian, 
             Cambridge, Massachusetts 02138, USA}

\author{H. R. Sadeghpour\,\orcidlink{0000-0001-5707-8675}}
\affiliation{ITAMP, Center for Astrophysics $|$ Harvard \& Smithsonian, 
             Cambridge, Massachusetts 02138, USA}


\begin{abstract}

Recent advances in polaritonic chemistry suggest that chemical reactions can be controlled via collective vibrational strong coupling (VSC) in a cavity. In this fully analytical work, we demonstrate that the collective vibrations of a molecular ensemble under VSC execute a beating with a period inversely proportional to the collective vacuum Rabi splitting. Significantly, this collective beating is imprinted on the local dynamics and resonantly suppresses individual molecular vibrations when a fraction of molecules are vibrationally excited, as in activated complexes formed in chemical reactions. This emergent beating occurs on significantly longer time scales than the individual molecular vibration or the cavity field oscillation period, peaking at the cavity-molecule resonance, consistent with polaritonic chemistry experiments. The cavity mediates an energy exchange between excited and ground-state molecules, affecting the dynamics of the entire ensemble. These findings suggest that the dynamics in polaritonic chemical reactions may not be in full equilibrium. In the ultra-strong coupling regime, we find that the local vibrations are modified by the cavity even at short time scales. Notably, these dynamical effects also extend to isotropic molecular ensembles in our model. Our analytical model offers insights into how collective VSC can dampen local molecular vibrations at resonance, potentially altering chemical reactivity.
\end{abstract}

\maketitle
\textit{Introduction.} Polaritonic chemistry has emerged as an appealing
branch of synthetic chemistry that promises mode selectivity and a cleaner approach to
controlling chemical kinetics~\cite{EbbesenRubioScholes, Ebbesen2016}.
Experiments have suggested the control of chemical reactions through collective
vibrational strong coupling (VSC) in ensembles of $N$ molecules ($N\approx 10^{6}\text{\,-\,}10^{12}$)
in microcavities, without external fields
~\cite{Hutchison2012, Hutchison2013, Lather2019, SimpkinsScience, EbbesenNMR, Murakoshi},
finding that optimal modification of reactivity occurs when
the cavity is in resonance with molecular vibrational modes
~\cite{Hutchison2012, Hutchison2013, Lather2019, SimpkinsScience, EbbesenNMR, Murakoshi}.

These experimental findings have led to an extensive body of theoretical and experimental
research, with the aim of gaining a mechanistic understanding of resonant VSC for
chemical reactivity
~\cite{HuoReview2023, JoelReview, Ruggenthaler2023, SimpkinsReview, HiraiReview, 
HutchinsonReview, JoelOpticalFilters, BorjessonReview, sidler2024spinglass}. 
Despite such efforts, the fundamental mechanisms behind the observed phenomena are poorly understood~\cite{EbbesenRubioScholes}, primarily because of the inherent
complexity and out-of-equilibrium nature of reaction dynamics.

One key challenge is to elucidate how a single-mode cavity can modify the local
molecular dynamics in a large molecular ensemble under collective resonant
VSC~\cite{EbbesenRubioScholes}. From an equilibrium perspective, for example
in transition state theory~\cite{MolecularDynamicsBook}, this seems counterintuitive
since in polaritonic systems, as e.g., described by the Tavis-Cummings
~\cite{TavisCummings} or the Hopfield model~\cite{Hopfield, Tutunnikov_2025},
there are two polaritonic modes and $N-1$ dark states
whose energies are unaffected by the cavity~\cite{JoelReview, HuoReview2023}.
The thermodynamic behavior at room temperature should therefore
be dominated by the dark states and not affected by the cavity
~\cite{JoelReview, HuoReview2023, LargeNproblem}.

In this work, we provide a fully analytical description of the out-of-equilibrium dynamics
of a molecular ensemble under VSC, by evolving fully or partially excited ensemble
configurations under collective coupling to a cavity. Our analytical solutions reveal the
emergence of a resonant collective beating~\cite{Tutunnikov_2025}, which imprints itself
on the local vibrational dynamics and resonantly suppresses individual molecular vibrations. The beating observed in our model arises from the interference between
two polariton frequencies, which emerge when molecular vibrations are resonantly coupled
to the cavity mode. This interference produces a fast and a slow oscillating mode, with
the slowly oscillating mode defining the beating frequency. 

The period of the emergent beating that suppresses the local molecular vibrations 
in an activated complex, is inversely proportional to the collective vacuum
Rabi splitting (VRS), and thus depends on the number of coupled molecules. 
These findings imply that (i) the VRS introduces a global dynamical timescale
that depends on the collective properties of the coupled system, rather than merely
on the timescales of the subsystems; and (ii) the dynamics of molecular ensembles
under VSC may not necessarily operate in equilibrium. 

The beating period exhibits a pronounced peak
around the cavity-molecule resonance, similar to the resonance dependence found
in polaritonic chemistry experiments
~\cite{Hutchison2012, Hutchison2013, Lather2019, SimpkinsScience, EbbesenNMR, Murakoshi}.
This phenomenon underlies the resonant suppression of local vibrations in our model.
It is worth noting that we describe the cavity-induced resonant beating in 
a classical framework, providing a clear intuition that resonant dynamical
modifications can be observed at the classical level. The quantum nature of
the observed phenomena in polaritonic chemistry remains an active research endeavor
~\cite{Li2021, Fiechter2023, JoelOpticalFilters, LindoyMandalReichman2024,
HuoQuantum2023, LindoyQuantum2023}. Recent works suggested that
the sharp resonance effect in polaritonic chemistry is purely quantum
~\cite{Li2021, Fiechter2023, LindoyMandalReichman2024, HuoQuantum2023, LindoyQuantum2023}.
Our findings within a classical formalism help shed light on the mechanisms
for resonant suppression and opens a path for understanding resonant collective
phenomena in polaritonic chemistry with classical dynamics simulations. It is important to note that the classical description of vibrational dynamics in our
model exactly reproduces the quantum dynamics of the first moments due to Ehrenfest’s
theorem for harmonic systems~\cite{ShankarBook, TannorBook}.

In the ultrastrong coupling regime, where the counter-rotating terms in the Hamiltonian are
significant~\cite{kockum2019ultrastrong, UltrastrongReview2019, QINultrastrong}, 
we find that the local molecular vibrations are modified by the cavity even
on a short time scale. In this case, the molecular vibrations are strongly
affected even during a single oscillation period, and the molecular dynamics
become more complicated and do not follow a simple beating pattern.  

Crucially, the cavity-induced modifications of the local vibrations require only
a partially activated molecular ensemble (where $\sim 1\text{\,-\,}5\%$ of the 
molecules are excited). This holds both in the strong and in the ultrastrong coupling regimes.
Additionally, in the partially excited ensemble, the initially excited
molecules transfer their energy to the ground-state ones. Thus, the cavity
mediates an energy exchange between the excited and the ground-state molecules. These important dynamical phenomena in our model can also be generalized
to the case of isotropic molecular ensembles within our model. The fact that full
activation or perfectly oriented molecules are not necessary, makes
the proposed mechanism relevant for realistic activated complexes, thus potentially
enabling alteration
and control of chemical reactivity.

\textit{Model Hamiltonian.} We consider a system of $N$ identical
non-interacting polar molecules collectively coupled
to a single-mode cavity. The molecular vibrations are modeled using a one-dimensional 
harmonic potential. For simplicity, the molecules are assumed to be perfectly oriented
along one of the cavity polarization directions. It is important to note,
however, that as we will show later, our model and analysis can also be generalized for
the case of an isotropic ensemble where half of the molecules are oriented along the 
field polarization and half are oriented against the field. This system is described by
the Pauli-Fierz Hamiltonian in the length gauge $(\hbar=1)$
~\cite{CohenTBook1997, rokaj2017, HuoReview2023},
\begin{equation}
\hat{\mathcal{H}}
\!=
\hat{\mathcal{H}}_0 -\frac{\omega \partial^2_q}{2}
+
\frac{\omega}{2}\!\left(\!\hat{q} \!-\! g \! \sum_{i=1}^N \hat{x}_i\right)^2\!.
\end{equation}
In the above, the Hamiltonian
$\hat{\mathcal{H}}_0 = \sum_{i=1}^N[-\partial^2_{x_i}/(2m) + m\Omega_\rmv^2\hat{x}_i^2/2]$
describes the uncoupled $N$ molecules, with $\Omega_\rmv$ being the equilibrium frequency
of the harmonic molecular potential, and $m$ is the molecular mass. The operators 
$\hat{x}_i$ and $-\mathrm{i}\partial_{x_i}$ correspond to the coordinates (bond lengths)  and
momenta of the molecules.  In addition, $\omega=\pi c / L$ is the fundamental frequency
of the cavity, and $c$ is the speed of light in vacuum. Here, we assume a standard
Fabry-P\'{e}rot cavity with effective optical volume $\mathcal{V}=\mathcal{A}L$, where
$\mathcal{A}$ is the effective cross-sectional area of the confinement volume, and $L$ is
the distance between the cavity mirrors~\cite{JoelReview}. The dimensionless light-molecule
coupling constant is $g=\mu_0 /\sqrt{\omega\epsilon_0\mathcal{V}}$, depending on 
$\mathcal{V}$, the vacuum permittivity $\epsilon_0$, and the magnitude of the molecular
dipole moment $\mu_0$ in units of elementary charge $e$. The operators $\hat{q}$ and 
$-\mathrm{i}\partial_{q}$ describe the position and momentum quadratures of the cavity mode
~\cite{HuoReview2023, rokaj2017}.

The model Hamiltonian $\hat{\mathcal{H}}$ has been used in multiple publications considering
molecular systems under VSC in cavities
~\cite{Yang2021, JoelReview, HuoReview2023, horak2024, JoelPolCond}.
The oriented molecules couple to the cavity through the collective dipole moment
operator $\hat{\mu}=\mu_0\sum_i\hat{x}_i$. This suggests that only one collective degree
of freedom is coupled to the cavity. To show this explicitly, we express $\hat{\mathcal{H}}$ 
in terms of the collective coordinate $\hat{X} = N^{-1/2} \sum_i \hat{x}_i$ and the relative 
bond lengths $\hat{\tilde{x}}_j = N^{-1/2}(\hat{x}_1-\hat{x}_j)$ with $j=2,\dots,N$. 
The prefactor $N^{-1/2}$ is introduced for mathematical convenience as in 
Refs.~\cite{BuschSol, RokajTopo2023}. The operators $\hat{X}$ and $\hat{\tilde{x}}_j$,
along with their corresponding momenta satisfy canonical commutation
relations as it was shown for the two-particle case in Ref.~\cite{BuschSol} and the $N$
particle case in Ref.~\cite{RokajTopo2023}.

In terms of new coordinates, the Hamiltonian is a sum of two independent
parts, $\hat{\mathcal{H}}=\hat{\mathcal{H}}_\mathrm{col}+\hat{\mathcal{H}}_\mathrm{rel}$,
where~\cite{Tutunnikov_2025, Vasilis2023}
\begin{equation} \label{eq:Hcmframe}
\begin{aligned}
\hat{\mathcal{H}}_\mathrm{col} 
& \!= 
-\frac{\partial^2_X}{2m}+\frac{m\Omega_\rmv^2}{2}\hat{X}^2 - \frac{\omega \partial^2_q}{2}
+\frac{\omega}{2} \! \left(\hat{q}-g\sqrt{N}\hat{X}\right)^2\!\!,\\
\hat{\mathcal{H}}_\mathrm{rel} & = \sum_{j=2}^N\left[-\frac{\partial^2_{\tilde{x}_j}}{2mN} 
+
\frac{Nm\Omega_\rmv^2}{2}\hat{\tilde{x}}_j^2 \right]  \\
&-\sum_{j,k=2}^N \frac{\partial_{\tilde{x}_j} \partial_{\tilde{x}_k}}{2mN}
-
\frac{m\Omega_\rmv^2}{2}\left[\sum_{j=2}^{N}\hat{\tilde{x}}_j\right]^2.
\end{aligned}
\end{equation}
The Hamiltonian $\hat{\mathcal{H}}_\textrm{col}$ describes the collective vibration coupled
to the cavity field, while $\hat{\mathcal{H}}_\textrm{rel}$ defines the dynamics of the
relative bond lengths, decoupled from the cavity and $\hat{X}$. For what follows, the relations
between the original operators $\hat{x}_i$'s and the new ones, $\hat{X}$ and
$\hat{\tilde{x}}_j$'s, will prove important,
\begin{equation} \label{eq:coordinates}
\begin{aligned}
\hat{x}_1 &= \frac{1}{\sqrt{N}}\left(\hat{X}+\sum^N_{j=2}\hat{\tilde{x}}_j\right),\\
\hat{x}_i &= \frac{1}{\sqrt{N}}\left(\hat{X}+\sum^N_{j=2}\hat{\tilde{x}}_j\right) 
           - \sqrt{N}\hat{\tilde{x}}_i;\;\;i=2,\ldots,N.
\end{aligned}
\end{equation}

\textit{Analytical Solution of the Hybrid System}. After expanding the
quadratic term $(\hat{q}-g\sqrt{N}\hat{X})^2$ in Eq.~\eqref{eq:coordinates} and introducing
the dressed vibrational frequency $\bar{\Omega}_\rmv^2 = \Omega_\rmv^2 + \omega^2_d$, where
$\omega_d=\sqrt{\mu^2_0N/(m\epsilon_0\mathcal{V})}$ is the so-called diamagnetic frequency 
or depolarization  shift~\cite{Todorov2010, Rokaj2deg}, $\hat{\mathcal{H}}_\mathrm{col}$
takes the form
$\hat{\mathcal{H}}_\mathrm{col} = -\partial^2_X/(2m) + m\bar {\Omega}_\rmv^2\hat{X}^2/2 
                                   - \omega \partial^2_q/2 + \omega \hat{q}^2/2 
                                   -\omega g\sqrt{N}\hat{q}\hat{X}$. 
Then, 
$\hat{\mathcal{H}}_\mathrm{col}$ can be brought into canonical form through the
transformation,
\begin{equation} \label{eq:polcor}
\begin{aligned}
\hat{X} &= \frac{1}{\sqrt{m}} \left(- \frac{\Lambda \hat{Q}_+}{\sqrt{1+\Lambda^2}} 
                                       + \frac{\hat{Q}_-}{\sqrt{1+\Lambda^2}}\right),\\
\hat{q} &= -\sqrt{\omega} \left(\frac{\hat{Q}_+}{\sqrt{1+\Lambda^2}} 
                               +\frac{\Lambda \hat{Q}_-}{\sqrt{1+\Lambda^2}}\right).
\end{aligned}
\end{equation}
Here, the parameter $\Lambda=\alpha-\sqrt{1+\alpha^2}$, with
$\alpha = (\omega^2 - \bar{\Omega}^2_\rmv)/(2 \omega_d \omega)$, quantifies the degree of
mixing between the cavity and the collective molecular degrees of freedom.

After the transformation, $\hat{\mathcal{H}}_\mathrm{col}$ takes the canonical form
$\hat{\mathcal{H}}_\mathrm{col} = \sum_{l=\pm}(-\partial^2_{Q_l}/2 + \Omega^2_l\hat{Q}^2_l/2)$.
The normal mode frequencies (polariton frequencies) of $\hat{\mathcal{H}}_\mathrm{col}$ are
\begin{equation}
\Omega^2_{\pm} = \frac{\bar{\Omega}^2_\rmv + \omega^2}{2}
\pm
\frac{1}{2}\sqrt{4\omega^2_d\omega^2 + (\bar{\Omega}^2_\rmv - \omega^2)^2}.
\end{equation} 
The modes $\Omega_+$ and $\Omega_-$ are the upper and lower polaritons, respectively. 
The difference between the polariton frequencies $\Delta=\Omega_+-\Omega_-$ is the 
polariton gap. At resonance, $\omega=\Omega_\rmv$, the polariton gap is equal to the vacuum
Rabi splitting (VRS)
$\Omega_{R}=\Delta_{\omega=\Omega_\rmv}=(\Omega_+-\Omega_-)_{\omega=\Omega_\rmv} 
= \sqrt{\omega_d^2(4\omega^2+\omega_d^2)}$, the fundamental spectroscopic observable
in hybrid light-matter systems. The normalized VRS, $\Omega_R/\Omega_\rmv$ defines the 
regime of light-matter interaction. Typically, $\Omega_R/\Omega_\rmv< 0.1$ is defined as
the strong coupling regime, while $0.1\leq\Omega_R/\Omega_\rmv< 1$ is the ultrastrong
coupling regime, where interaction terms beyond the rotating-wave approximation become
important~\cite{kockum2019ultrastrong, UltrastrongReview2019}.

From the canonical form of $\hat{\mathcal{H}}_\mathrm{col}$ follows the standard
time-dependent solution for the polariton operators,
\begin{equation} \label{eq:polsolution}
\hat{Q}_\pm (t) = \hat{Q}_\pm (0) \cos(\Omega_\pm t) 
+ 
\frac{\hat{\dot{Q}}_\pm (0)}{\Omega_\pm} \sin(\Omega_\pm t), 
\end{equation}
which allows us to obtain $\hat{X}(t)$ and $\hat{q}(t)$ via Eq.~\eqref{eq:polcor}.

Next, we focus on the relative bond lengths $\hat{\tilde{x}}_j$'s described by
$\hat{\mathcal{H}}_\mathrm{rel}$. A crucial observation is that the cavity mode is absent in
$\hat{\mathcal{H}}_\mathrm{rel}$, indicating that the evolution of $\hat{\tilde{x}}_j$'s
remains unaffected by the cavity. Consequently, we can obtain the general time-dependent
solutions for $\hat{\tilde{x}}_j(t)$ simply by using the uncoupled Hamiltonian
$\hat{\mathcal{H}}_0 = \sum_{i=1}^N[-\partial^2_{x_i}/(2m) + m\Omega_\rmv^2\hat{x}_i^2/2]$.

The standard time-dependent solution for the position operators given $\hat{\mathcal{H}}_0$ is
$\hat{x}_i(t) = \hat{x}_i(0) \cos(\Omega_\rmv t) + \hat{p}_i(0)\sin(\Omega_\rmv t)/(m\Omega_\rmv)$.
Substitution into the definition,
$\hat{\tilde{x}}_{j} = N^{-1/2}(\hat{x}_1-\hat{x}_j)$, yields
\begin{equation} \label{eq:x_j_relative}
\hat{\tilde{x}}_{j}(t)
\!=\!
\frac{\hat{x}_1(0)\!-\!\hat{x}_j(0)}{\sqrt{N}} \!\cos(\Omega_\rmv t) 
\mkern-1mu +
\frac{\hat{p}_1(0)\!-\!\hat{p}_j(0)}{\sqrt{N}m\Omega_\rmv}\!\sin(\Omega_\rmv t), \! \!
\end{equation}
where $j=2,\dots,N$. This shows that all relative bond lengths $\hat{\tilde{x}}_{j}$
oscillate with the bare vibrational frequency $\Omega_\rmv$. Thus, we have found the full
eigenspectrum of $\hat{\mathcal{H}}$ describing the $N$ molecules coupled to the cavity,
$\Omega_-, \{\Omega_\rmv\}_{N-1}, \Omega_+$. The spectrum has two polariton modes
$\Omega_{\pm}$ and $N-1$ modes $\Omega_\rmv$ unaffected by the cavity, also known as dark
states. We note that the eigenmodes and dynamics of the hybrid system could
also be obtained by diagonalizing a Hessian matrix, as it was done in 
Ref.~\cite{ZhouPolaritonModes} in the rotating-wave approximation. Our solution reproduces
the polariton modes obtained in~\cite{ZhouPolaritonModes} under the rotating-wave approximation.

\textit{Dynamics of the First Moments}. According to the Ehrenfest theorem,
the first moments (e.g., $\braket{\hat{X}(t)}$, $\braket{\hat{q}(t)}$, 
$\braket{\hat{x}_i(t)}$, etc.) exactly equal the corresponding classical solutions in 
a harmonic system~\cite{ShankarBook, TannorBook}. Position-momentum
uncertainty manifests itself only in higher-order observables. Thus, we employ classical
mechanics to provide a clear description of the physical phenomena in our system at the
level of the first moments under different initial conditions. With this in mind, we omit
the operator hats for convenience from this point on.

To find $X(t)$ and $q(t)$, we substitute $Q_{\pm}(t)=A_{\pm}\sin(\Omega_{\pm}t+\phi_{\pm})$
[Eq.~\eqref{eq:polsolution}] into Eq.~\eqref{eq:polcor},
\begin{equation} \label{eq:collective_sols}
\begin{aligned}
X(t) \! &= \! -\frac{\Lambda A_+ \sin(\Omega_+ t + \!\phi_+)}{\sqrt{m(1+\Lambda^2)}}
     \! +  \!
               \frac{A_{-}\sin(\Omega_- t + \!\phi_-)}{\sqrt{m(1+\Lambda^2)}},\\
\frac{q(t)}{-\sqrt{\omega}} &= \frac{A_+ \sin(\Omega_+ t + \phi_+)}{\sqrt{1+\Lambda^2}}
    \!  +  \!
    \frac{\Lambda A_- \sin(\Omega_- t + \phi_-)}{\sqrt{1+\Lambda^2}},
\end{aligned}
\end{equation}
where the amplitudes $A_{\pm}$ and phases $\phi_{\pm}$ are determined by the initial
conditions of the cavity-molecules system.

Next, from Eq.~\eqref{eq:x_j_relative} we find the relative bond lengths,
$\tilde{x}_j(t)=A_j\sin\left(\Omega_\rmv t+ \phi_j\right)$ ($j=2,\dots,N$), where $A_j$
and $\phi_j$ are determined by the initial conditions. By substituting $X(t)$ and
$\tilde{x}_j(t)$ into Eq.~\eqref{eq:coordinates} we can obtain the time-dependent bond
lengths of each molecule, $x_i(t)$.

Having found the complete solution for the
molecular ensemble coupled to the cavity, we will investigate the dynamics of collective
and local molecular vibrations under different initial non-equilibrium conditions. 
Such initial conditions are motivated by an attempt to provide insights into
cavity-modified chemical reactivity repeatedly reported in the literature
~\cite{Hutchison2012, Hutchison2013, Lather2019, SimpkinsScience, EbbesenNMR, Murakoshi}.
Out-of-equilibrium states naturally emerge within activated chemical complexes
~\cite{MolecularDynamicsBook}. 
The activated complex comprises vibrationally active molecules (reactants) that
approach each other and realize a transient intermediate state, where one or more 
molecular bonds are stretched, compressed, or distorted~\cite{MolecularDynamicsBook}.
The displaced bonds can be thought of as harmonic oscillators displaced from equilibrium.
The vibrational modes of the activated complex often dominate the reaction dynamics,
making them critical degrees of freedom for reaction kinetics.
The notion of an activated complex is crucial for chemical reactivity and plays
an important role in the transition-state theory~\cite{MolecularDynamicsBook}. However, we would like to clarify that with our model we do not try to match the initial conditions in transition-state theory, which operates in pure thermal equilibrium. In addition, transient atomic arrangements are important in roaming reactions
~\cite{BowmanScience, RoamingReactionReview}. 
In what follows, we focus on the transient, out-of-equilibrium behavior of an ensemble of molecules under
collective VSC in a cavity.

\textit{Cavity-Induced Resonant Collective Beatings.} First, we look at the case of a fully excited ensemble, where all molecular bonds
are initially stretched by $x_0\ll1$, i.e. $x_i(0)\approx x_0$, while the initial velocities,
$\dot{x}_i(0)=v_i$ satisfy the condition $\sum_i v_i=0$, such that the velocities of
individual molecules may not be zero, but the average vanishes. The cavity
mode is considered to be unperturbed, i.e. $q(0)=\dot{q}(0)=0$. This is motivated by
the polaritonic chemistry experiments with vacuum cavities, i.e., without external 
fields~\cite{Hutchison2012, Hutchison2013, Lather2019, SimpkinsScience, EbbesenNMR, Murakoshi}.
\begin{figure}
\includegraphics[width=\linewidth]{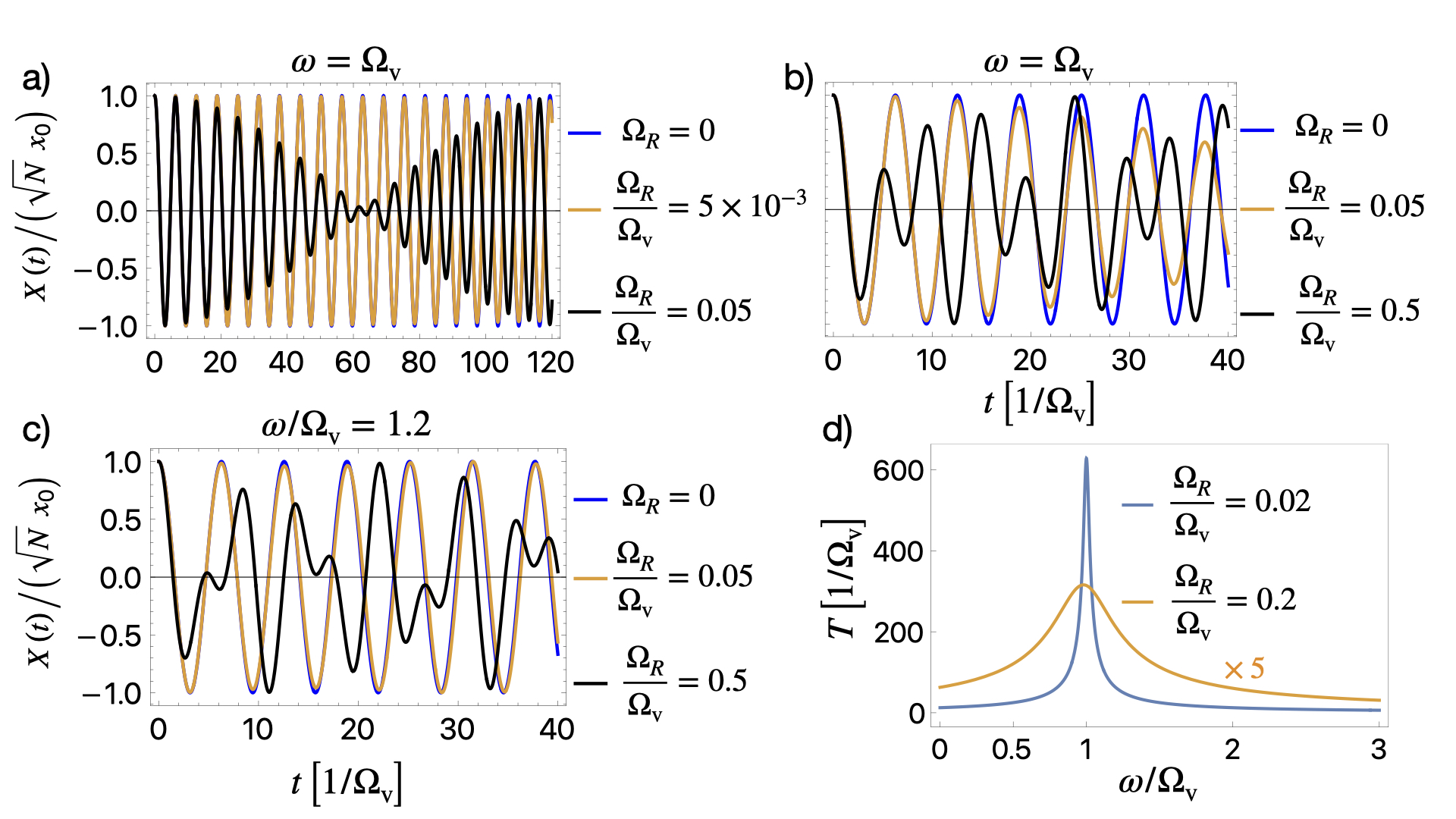}
    \caption{Evolution of the collective molecular coordinate, $X(t)$
    [see Eqs.~\eqref{eq:collective_sol} and \eqref{eq:collective_sol_strong}] for
    different values of the coupling constant. 
    (a) For resonant strong coupling ($\omega=\Omega_\rmv$, $\Omega_R/\Omega_\rmv=0.05$,
    black), slow beatings emerge. For weaker coupling ($\Omega_R/\Omega_\rmv=0.005$), 
    the beating effect is negligible, and becomes apparent only on extremely long
    time scales.
    (b) For ultrastrong coupling ($\Omega_R/\Omega_\rmv=0.5$, black) the cavity affects
    strongly also the short-time dynamics, as the collective vibration is suppressed during
    the first period of oscillation. For strong coupling, there is no visible effect
    on the short time scale.
    (c) Out of resonance ($\omega=1.2\Omega_\rmv$), the collective vibrations are modified
    for ultrastrong coupling ($\Omega_R/\Omega_\rmv=0.5$, black), while for strong coupling
    ($\Omega_R/\Omega_\rmv=0.05$, orange) they remain unaffected. 
    (d) Dependence of the beating period $T$ for strong ($\Omega_R/\Omega_\rmv=0.02$, blue) 
    and ultrastrong ($\Omega_R/\Omega_\rmv=0.2$, orange) couplings. For strong coupling, $T$ 
    peaks sharply around the resonance point, while for ultrastrong coupling, the beating 
    period broadens and becomes asymmetric with respect to the resonance point.}
\label{fig:collective_vib}
\end{figure}

Consequently, for the collective coordinate, we have $X(0)=\sqrt{N}x_0$ and $\dot{X}(0)=0$,
and in combination with the initial conditions of the cavity, we find the phases
$\phi_{\pm}=\pi/2$ and the amplitudes $A_{-}=x_0\sqrt{mN/(\Lambda^2+1)}$, 
$A_+=-\Lambda A_-$ such that the collective coordinate in Eq.~\eqref{eq:collective_sols}
reads
\begin{equation} \label{eq:collective_sol}
X(t) = x_0\sqrt{N} \left[\frac{\Lambda^2\cos(\Omega_+ t)}{\Lambda^2 +1}
                       + \frac{\cos(\Omega_- t)}{\Lambda^2 +1} \right].
\end{equation}
In the strong coupling regime (not ultrastrong coupling
~\cite{kockum2019ultrastrong, UltrastrongReview2019}), the renormalization of the bare
molecular vibrational frequency is negligible, i.e. $\bar{\Omega}_\rmv\approx \Omega_\rmv$.
Then, for the cavity resonantly coupled to the vibrations $\omega=\Omega_\rmv$, the mixing
parameter is $\Lambda=-1$ and the two polariton modes contribute equally to the collective
motion. Using the sum-to-product trigonometric identity, the collective vibration reads
\begin{equation} \label{eq:collective_sol_strong}
X(t) \!= x_0\sqrt{N}\cos\!\left(\frac{\Omega_+ + \Omega_- }{2}t\right)
                  \cos\!\left(\frac{\Omega_+ - \Omega_- }{2}t\right).
\end{equation}
Under VSC, the collective coordinate oscillates with two frequencies: the high
$\Sigma=\Omega_+ + \Omega_-$, and the low $\Delta=\Omega_+-\Omega_-$, which is equal to
the polariton gap. Due to the polariton gap, the collective vibration exhibits slow
beatings with a period $T=4\pi/ \Delta$. At resonance, the beatings are inversely
proportional to the VRS, $T|_{\omega=\Omega_\rmv}=4\pi/\Omega_R$. The VRS depends on
the number of molecules collectively coupled as $\Omega_R\sim \sqrt{N}$. Thus, the VSC
introduces an emerging time scale in the molecular dynamics, which depends on the collective
properties of the coupled system. We note that cavity-induced collective beatings
were first obtained with fully quantum mechanical simulations in Ref.~\cite{Tutunnikov_2025}. It is important to note, that the influence of the collective beatings to the local molecular dynamics was not studied in~\cite{Tutunnikov_2025}. The local vibrational dynamics are the main focus of this work as we will see in what comes later. 
\begin{figure*}
\includegraphics[width=0.8\linewidth]{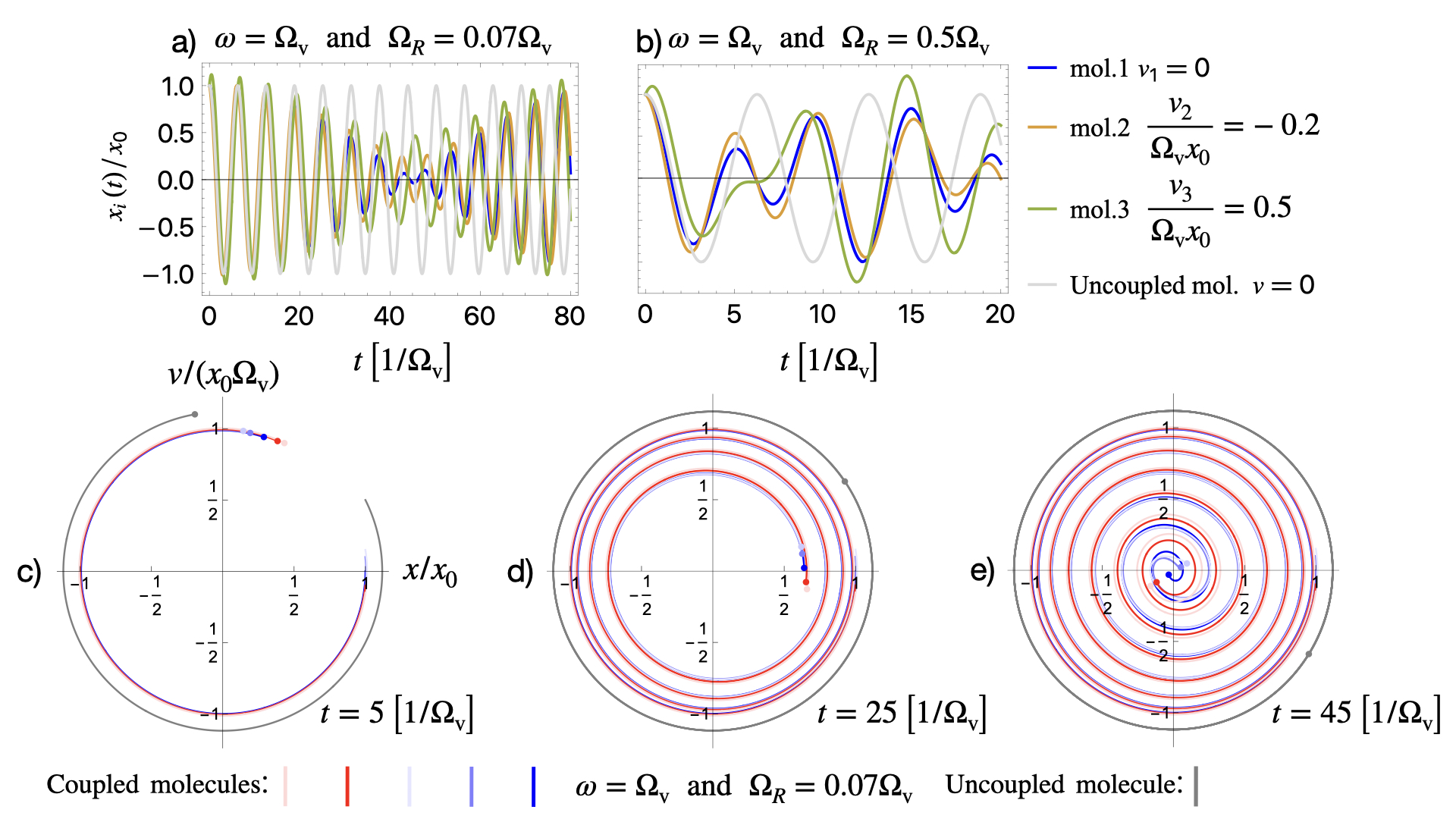}
\caption{(a,b) Evolution of the vibrations of three representative molecules 
    (out of the total of $N$ molecules in the ensemble) with different initial velocities
    under resonant $\omega=\Omega_\rmv$ and collective VSC characterized by the normalized VRS 
    $\Omega_R/\Omega_\rmv$.
    (a) For strong coupling $\Omega_R/\Omega_\rmv=0.07$, all molecules experience suppression
    of their local vibrations due to the cavity-induced resonant beatings at long time scales. 
    (b) For ultrastrong coupling $\Omega_R/\Omega_\rmv=0.5$, molecular vibrations are strongly
    modified at short times, even during the first oscillation.
    (c-e) Visualization of the phase space trajectories of local molecular vibrations
    for resonant coupling $\omega=\Omega_\rmv$ and normalized VRS $\Omega_R/\Omega_\rmv=0.07$ .
    (i) At short times, all molecules seem to follow a circular trajectory around the origin.
    (ii) At later times, however, the trajectories of coupled molecules spiral towards the 
    phase space origin as their vibrational energy is transferred to the cavity.
    (iii) The cavity-induced resonant beatings force the molecules to spend more time close to
    the phase-space origin, where their vibrational energy is low. This means that the local 
    molecular vibrations are effectively frozen during this time. For comparison, an uncoupled
    molecule (gray trajectory) follows a fixed circular trajectory with constant total energy.}
\label{fig:phase_space}
\end{figure*}

In Fig.~\ref{fig:collective_vib}, we show the evolution of the collective molecular
coordinate. In Fig.~\ref{fig:collective_vib}(a,b), we focus on resonant coupling,
$\omega=\Omega_\rmv$. For strong coupling in Fig.~\ref{fig:collective_vib}(a)
($\Omega_R/\Omega_\rmv=0.05$, black line), we observe the emergence of slow beatings
which significantly suppress collective vibrations at late times, while the short-time
dynamics modifications are insignificant. The observed beating arises from
the interference between two polariton frequencies, $\Omega_{\pm}$, which emerge when
molecular vibrations are resonantly coupled to the cavity mode. This interference produces
a fast and a slow oscillating modes, with the slowly oscillating mode defining the beating
frequency.

In the same panel, for weaker coupling
($\Omega_R/\Omega_\rmv=0.005$),
no visible modification is observed, as the beatings become apparent only after extremely
long times (not shown in the figure). For ultrastrong coupling ($\Omega_R/\Omega_\rmv=0.5$)
considered in Fig.~\ref{fig:collective_vib}(b), even the short-time dynamics are modified
by the cavity. This is a distinctive feature of ultrastrong coupling compared to strong. Next, in Fig.~\ref{fig:collective_vib}(c), we detune the cavity from the
vibrational mode by setting $\omega/\Omega_\rmv=1.2$ and observe that for strong coupling
($\Omega_R/\Omega_\rmv=0.05$, orange line) no modification is observed. This striking
finding highlights the importance of resonant coupling in modifying molecular
vibrations in the strong coupling regime. This behavior aligns with the experimental
observations in polaritonic chemistry, where the light-matter interaction is in the strong
coupling regime, and modifications of the chemical reactivity occur mainly on resonance
~\cite{Hutchison2012, Hutchison2013, Lather2019, SimpkinsScience, EbbesenNMR, Murakoshi}. 

Under ultrastrong coupling ($\Omega_R/\Omega_\rmv=0.5$, black line), on the other
hand, the collective vibrations are modified even off-resonance. Thus, we find that
ultrastrong coupling has two important advantages compared to strong coupling:
it allows for off-resonant and short-time modifications of the vibrational dynamics.
These important features are also visualized in Fig.~\ref{fig:collective_vib}(d) where
the beating period $T$ sharply peaks around the light-matter resonance for strong
coupling (blue line), while for ultrastrong coupling, the curve broadens significantly. Thus, under ultrastrong coupling, molecular vibrations can be modified
off-resonance as the beating period broadens around the cavity-molecule
resonance (see Fig.~\ref{fig:collective_vib}). This phenomenon could inspire new
experimental approaches and deepen our understanding of vibrational dynamics in cavities.

Before continuing, we should discuss the values of the relevant experimental parameters
required to observe the phenomena shown in Fig.~\ref{fig:collective_vib} in the strong 
and ultrastrong coupling regimes. Assuming a terahertz Fabry-P\'{e}rot cavity with
frequency $\omega=2\pi \times 1 \, \rm{THz}$ and effective cross-sectional area
$\mathcal{A}=1 \, \upmu \mathrm{m}^2$. The molecular mass is assumed to be
$m=4\times 10^3 \, m_e$, where $m_e$ is the electron mass, the magnitude of
the molecular dipole moment for simplicity is assumed to be one in units of
fundamental charge $\mu_0=1e$, and the molecular vibration
frequency is equal to the cavity mode $\Omega_\rmv=\omega=2\pi \times 1\,\rm{THz}$.
Given these parameters to reach collective strong coupling $\Omega_{R}/\Omega_\rmv=0.02$
the number of collectively coupled molecules needs to be $N\approx 10^7$, while for
ultrastrong coupling $\Omega_{R}/\Omega_\rmv=0.2$ we need $N\approx 10^9$.
Thus, the cavity-induced collective beatings we uncover occur for the relevant parameters
typically reported in polaritonic chemistry experiments
~\cite{HuoReview2023, JoelReview, Ruggenthaler2023}. 
The same range of parameters will be used in what follows, where we focus on
local vibrational dynamics.

\textit{Resonant Suppression of Local Molecular Vibrations.} A key challenge in polaritonic chemistry is to understand whether cavity-induced
modifications of collective molecular dynamics indeed affect the chemistry at the level
of individual molecules. To address this, we investigate the evolution of the
relative bond lengths $\{\tilde{x}_j(t)\}_{j=2,\dots,N}$ in order to access the individual
local vibrations under the same initial conditions as previously, i.e.,
$x_i(0)\approx x_0$ for $i=1,\dots,N$, $\sum_iv_i=0$, and $q(0)=\dot{q}(0)=0$.
In this case, the initial relative bond lengths and velocities are all zero
$\tilde{x}_j(0)=0$, and thus the phases are all zero $\phi_j=0$ for $j>1$.
The initial velocities of the relative coordinates are
$\tilde{v}_j(0)=N^{-1/2}(v_1-v_j)$. Without loss of generality, we assume $v_1=0$ and we
have $\tilde{v}_j(0)=-v_j/\sqrt{N}\; \textrm{with}\; j>1$. Thus, we find
$\tilde{x}_j(t)=-v_j \Omega_\rmv \sin(\Omega_\rmv t)/\sqrt{N}$ for $j>1$. 

Since the sum over the initial velocities is zero, using
Eqs.~\eqref{eq:coordinates} and \eqref{eq:collective_sol}, we find the solutions
for all molecules ($i=1,\dots,N$),
\begin{equation} \label{eq:local_solutions}
x_i(t) \!=\! \underbrace{x_0\!\left[\frac{\Lambda^2\cos(\Omega_+ t)}{\Lambda^2 +1}
\!+\! 
\frac{\cos(\Omega_- t)}{\Lambda^2 +1} \right]}_\textrm{polaritonic}
\!+ \underbrace{\frac{v_i}{\Omega_\rmv}\sin(\Omega_\rmv t)}_\textrm{bare molecular}.\!\!
\end{equation}
Thus, vibrations of individual molecules are superpositions of vibration with bare molecular
frequency $\Omega_\rmv$ and collective molecular vibration, defined by polariton
frequencies, $\Omega_{\pm}$. Importantly, the polaritonic contribution has no prefactor $1/N$,
which would make it negligible at the limit of $N\rightarrow\infty$. This is one of the key
findings of this work, showing that for non-equilibrium initial conditions, the polaritons
modify dramatically the local molecular vibrations. Note that $\Lambda$ depends on
the particle density $N/\mathcal{V}$ only via $\omega_d=\sqrt{\mu^2_0N/(m\epsilon_0\mathcal{V})}$
which attains a finite value in the large $N$ (thermodynamic) limit. The only situation in
which the molecules are not affected by the polaritonic modes is if their initial vibrational
velocities are so large that $v_i\gg x_0 \Omega_\rmv$.

In Fig.~\ref{fig:phase_space}, we visualize the local molecular dynamics under resonant VSC
for different values of the VRS and different initial velocities.
Fig.~\ref{fig:phase_space}(a) shows the emergence of beatings in local vibrations under
strong coupling ($\Omega_R/\Omega_\rmv=0.07$) that suppresses molecular vibrations on
the long time scale. For ultrastrong coupling ($\Omega_R/\Omega_\rmv=0.5$) in
Fig.~\ref{fig:phase_space}(b), we observe that the short-time dynamics are also
significantly affected by the collective oscillations. Next, in
Figs.~\ref{fig:phase_space}(c-e) we provide the phase space visualization of the local
vibrations for multiple molecules with different initial velocities and normalized VRS
$\Omega_R/\Omega_\rmv=0.07$. At short times [$t=5/\Omega_\rmv$, 
Fig.~\ref{fig:phase_space}(c)] all molecules seem to follow a circular trajectory around
the origin. However, with time [$t=25/\Omega_\rmv$, Fig.~\ref{fig:phase_space}(d)],
the trajectories gradually spiral towards the origin. This indicates that the vibrational energy
is being transferred to the cavity. The cavity-induced resonant beatings force the
molecular trajectories to spend more time closer to the phase-space origin, where their energy
is low. This means that local molecular vibrations are effectively temporarily frozen 
[see Fig.~\ref{fig:phase_space}(e)]. For comparison, an uncoupled molecule [gray
trajectories in Figs.~\ref{fig:phase_space}(c-e)] follows a fixed circular trajectory with
constant total energy. Thus, from Fig.~\ref{fig:phase_space} it becomes clear that the
modifications of the collective vibrations $X(t)$ are imprinted on the local molecular
dynamics and drastically modify the individual molecular vibrations. This suggests a unique and general mechanism for modifying local
molecular vibrations under resonant collective VSC in a cavity, particularly in the
limit of many molecules. Thus we expect that the cavity-induced resonant beating, can
herald significant modifications in in the out-of-equilibrium vibrational dynamics of
molecules in activated complexes formed during chemical reactions.

\textit{Modification of Local Vibrations in a Partially Activated Ensemble.}
So far, we have considered the impact of cavity coupling on the vibrational dynamics of a
fully excited molecular ensemble. However, in a more realistic scenario, only a fraction
of the molecules are usually excited. Thus, we assume that there are $N_{\rm{ex}}$
excited molecules or, equivalently, a fraction $\beta=N_{\rm{ex}}/N$ of the ensemble is
displaced from equilibrium. For simplicity, we assume the excited molecules to be
approximately stretched by the same small displacement 
$x_i(0)\approx x_0, \; \forall \; i \in \{1,\dots,N_{\rm{ex}}\}$ while $x_{i^{\prime}}(0)=0$
for $i^{\prime} \neq i$. In addition, we assume that all initial velocities are zero
$\dot{x}_l(0)=v_l=0 \; \forall\; l=1,\dots,N$. As previously, the cavity is not initially
excited $q(0)=\dot{q}(0)=0$.

Correspondingly, the initial conditions for the collective vibration are
$X(0)=\beta \sqrt{N}x_0$ and $\dot{X}(0)=0$. Combining them with the cavity initial
conditions, we find the phases $\phi_{\pm}=\pi/2$ and the amplitudes
$A_{-}=\beta x_0\sqrt{mN/(\Lambda^2+1)}$, $A_+=-\Lambda A_-$ which determine the dynamical
evolutions in Eq.~\eqref{eq:collective_sols}. For the collective vibrations, we find
\begin{equation} \label{eq:collective_complex}
X(t) = \beta x_0\sqrt{N} 
       \left[\frac{\Lambda^2\cos(\Omega_+ t)}{\Lambda^2 +1} 
             + \frac{\cos(\Omega_- t)}{\Lambda^2 +1} \right].
\end{equation}
For finite $\beta$, the collective vibrations exhibit
beatings shown in Fig.~\ref{fig:collective_vib}(a) with $\beta$ merely controlling the
vibration amplitude. For zero initial velocities, the solutions for the relative 
bond length are $\tilde{x}_j(t)=N^{-1/2}[x_0-x_j(0)]\cos(\Omega_\rmv t)$. Substituting the
solutions for the collective vibration $X(t)$ and the relative bond lengths
$\tilde{x}_j(t)$ into Eq.~\eqref{eq:coordinates}, we obtain the solutions for the excited
molecules,
\begin{equation} \label{activated}
\frac{x_i(t)}{x_0}
\!\!=\!
\frac{\beta\Lambda^2\cos(\Omega_+ t)}{\Lambda^2 +1} 
+
\frac{\beta\cos(\Omega_- t)}{\Lambda^2 +1} + (1 \!-\! \beta)\cos(\Omega_\rmv t),\!\!
\end{equation}
where $i\in \{1,\dots,N_{\rm{ex}}\}$. For the ground-state molecules in the ensemble
\begin{equation}\label{nonactivated}
\frac{x_{i^{\prime}}(t)}{x_0}
=
\frac{\beta\Lambda^2\cos(\Omega_+ t)}{\Lambda^2 +1} 
+
\frac{\beta\cos(\Omega_- t)}{\Lambda^2 +1} -\beta\cos(\Omega_\rmv t),
\end{equation}
where $i^{\prime}\neq i$. From the above equations, we see that the local vibrations of
both excited and ground-state molecules are affected by the polaritonic modes
$\Omega_{\pm}$ that emerge due to the collective vibration $X(t)$
[see Eq.~\eqref{eq:collective_complex}]. At the same time, the local vibrations depend on
the bare molecular frequency $\Omega_\rmv$ as expected. 

\begin{figure}
\includegraphics[width=0.85\linewidth]{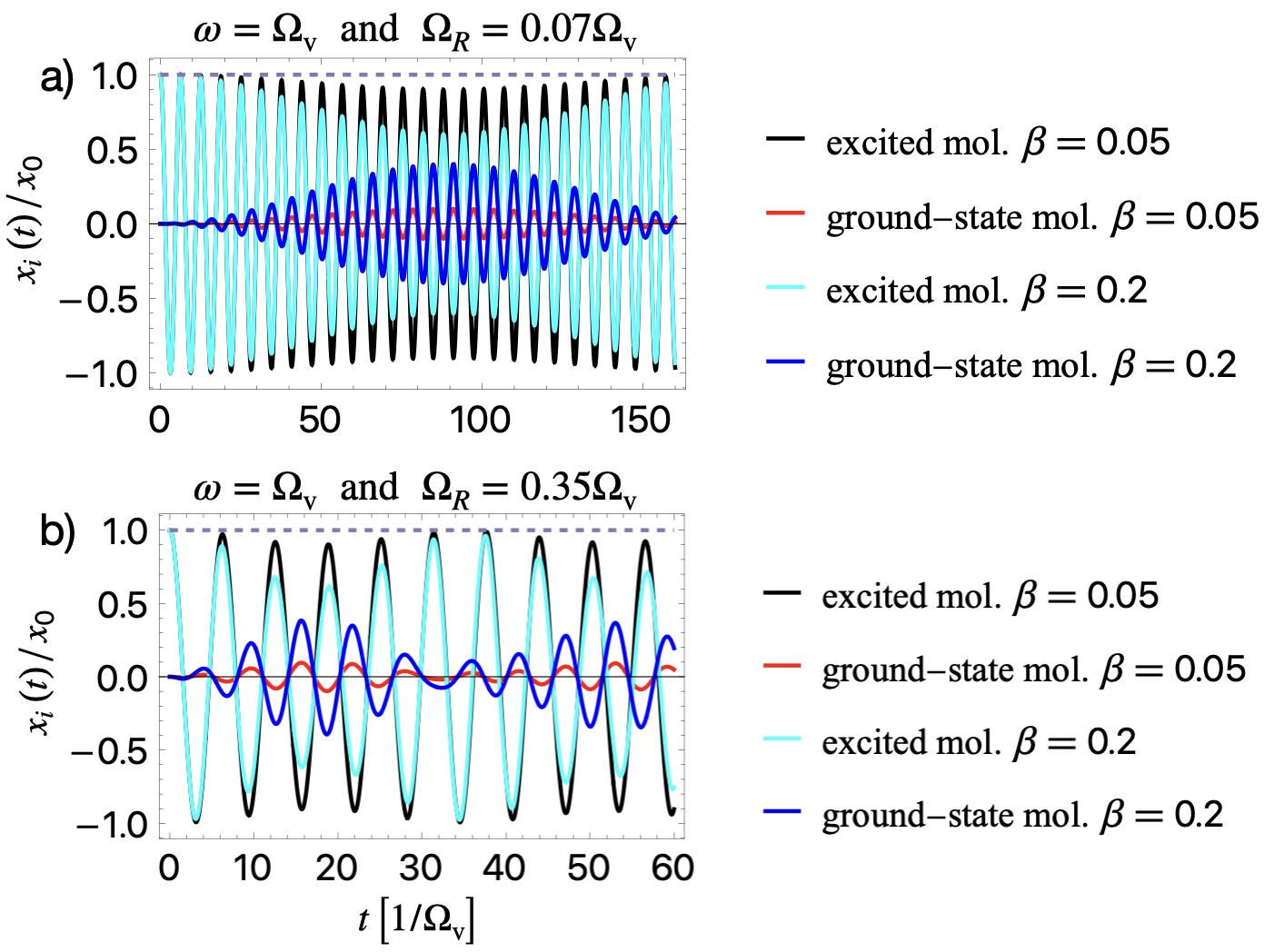}
    \caption{Evolution of the local molecular vibrations in a partially activated
    ensemble. 
    (a) For resonant strong coupling, $\omega=\Omega_\rmv$, $\Omega_R/\Omega_\rmv=0.07$,
    we observe that the excited molecules experience cavity-induced slow beatings,
    which suppress their vibrations (black and cyan). Conversely, the ground-state
    (red and blue) molecules begin vibrating with time and also
    exhibit beatings due to cavity-mediated energy exchange with originally excited molecules. 
    (b) For ultrastrong coupling, $\Omega_R/\Omega_\rmv=0.35$, the same dynamical behaviors
    as in (a) occur, but with a faster rate, as the beatings in excited and ground-state
    molecules appear at much earlier times. The amplitude of the beating depends linearly
    on the activation ratio $\beta=N_{\textrm{ex}}/N$.}
    \label{activated_complex}
\end{figure}

In Fig.~\ref{activated_complex}, we show the time evolution for the local vibrations of
excited and ground-state molecules resonantly coupled to the cavity $\omega=\Omega_\rmv$,
for different values of the activation ratio $\beta=0.05,\,0.2$ and
different values of the collective VRS. For strong cavity-molecule coupling 
$\Omega_R=0.07\Omega_\rmv$ in Fig.~\ref{activated_complex}(a), we observe that excited
(initially stretched) molecules experience cavity-induced slow beatings,
which substantially suppress their vibrations (black and cyan). Conversely, the
initially not stretched molecules (blue and red) begin vibrating with time and exhibit
beatings due to energy exchange with originally excited molecules. Thus, the cavity mediates
this energy transfer, which is optimal under resonant coupling. 
This phenomenon appears for both $5\%$ activation in the ensemble ($\beta=0.05$)
and for $20\%$ ($\beta=0.2$), which is more pronounced for the higher activation ratio
$\beta=0.2$, as intuitively expected. It is important to note that when the ensemble is
partially vibrationally excited, the cavity mediates intermolecular energy exchange.
In contrast, this phenomenon does not occur in a fully excited ensemble.
This highlights the importance of the molecular initial states for the cavity-mediated
dynamical correlation effects between the molecules in the ensemble.

In Fig.~\ref{activated_complex}(b), for ultrastrong coupling $\Omega_R=0.35\Omega_\rmv$ the
same phenomenon occurs but with a faster rate, as the beatings in the excited
and ground-state molecules appear at much earlier times. 
Overall, Fig.~\ref{activated_complex} demonstrates that the local vibrations of molecules
in a partially activated ensemble are strongly affected under resonant VSC to a cavity.
This is a significant result that broadens our understanding of the mechanisms that can
affect local vibrations and reactivity in polaritonic chemistry. These changes are driven by strong cavity-molecule coupling,
resulting in polariton formation, which highlights the intricate interplay between excited
molecular ensembles and the cavity environment.

It is important to note that in the absence of excited molecules ($N_{\rm{ex}}=0$), only
the solution in Eq.~\eqref{nonactivated} applies, and for $\beta=0$ becomes trivial
$x_{i^{\prime}}(t)=0$ meaning that we have no local vibrations as expected. In the opposite
case of the fully activated ensemble ($N_{\rm{ex}}=N$), only the solution in
Eq.~\eqref{activated} is true, and for $\beta=1$ we recover the result in
Eq.~\eqref{eq:local_solutions} for the case of zero initial velocities $v_i=0$. 
This demonstrates the consistency of our solution.

\textit{Isotropic Ensemble.} So far in this work, we considered the simple
case of a fully oriented ensemble where all the molecules are oriented along the cavity
field polarization. In our one-dimensional model, an approximate representation of an isotropic ensemble which has zero total dipole, can be considered by having half of the molecular dipoles in the direction of the field polarization (oriented) and half of the dipoles in the opposite direction (anti-oriented). Thus, the Hamiltonian takes the form 
\begin{eqnarray}
    \hat{\mathcal{H}}_{\rm iso} 
    &=& 
    \sum_{i=1}^N\left[-\frac{1}{2m}\frac{\partial^2}{\partial x^2_{i}} 
    \!+\!
    \frac{m\Omega^2_\rmv}{2} \hat{x}_i^2
    \!-\!
    \frac{1}{2m}\frac{\partial^2}{\partial s^2_{i}} 
    \!+\! 
    \frac{m\Omega^2_\rmv}{2} \hat{s}_i^2\right] \nonumber \\    
    &-& \frac{\omega \partial^2_q}{2}
    \!+\!
    \frac{\omega}{2}\!\left(\!\hat{q} 
    \!-\! 
    g  \! \sum_{i=1}^N \hat{x}_i 
    \!+\! 
    g \sum_{i=1}^N \hat{s}_i \right)^2\!.
\end{eqnarray}
The operators $\{\hat{x}_i,\partial_{x_i}\}$ describe the molecules oriented along the field, while the operators $\{\hat{s}_i,\partial_{s_i}\}$ describe the anti-oriented ones. 
This can be understood from the opposite sign in their respective light-matter couplings.
Note that all molecules have the same mass $m$ and the same vibrational frequency $\Omega_\rmv$.
The cavity mode couples only to the collective coordinates of the aligned and the anti-aligned
molecules respectively 
$\hat{X} = N^{-1/2} \sum_i \hat{x}_i$ and $\hat{S} = N^{-1/2} \sum_i \hat{s}_i$, while the relative 
bond lengths 
$\hat{\tilde{x}}_j = N^{-1/2}(\hat{x}_1-\hat{x}_j)$ and $\hat{\tilde{s}}_j = N^{-1/2}(\hat{s}_1-\hat{s}_j)$
with $j=2,\dots,N$ decouple from the cavity mode. This can be seen straightforwardly in the
new coordinate frame,
\begin{eqnarray}
    \hat{\mathcal{H}}_{\rm{iso}} &=&
    - \frac{\partial^2_X}{2m}+\frac{m\Omega_{\rm{v}}^2}{2}\hat{X}^2 -\frac{\partial^2_S}{2m}
    + \frac{m\Omega_{\rm{v}}^2}{2}\hat{S}^2  - \frac{\omega \partial^2_q}{2}\nonumber \\
    &+& \frac{\omega}{2} \! \left(\hat{q}-g\sqrt{N}\hat{X}+ g\sqrt{N}\hat{S}\right)^2 
    + \hat{\mathcal{H}}_{\rm{rel}}(\hat{\tilde{x}}_j , \hat{\tilde{s}}_j ).
\end{eqnarray}
In the above, the term $\hat{\mathcal{H}}_{\rm{rel}}(\hat{\tilde{x}}_j, \hat{\tilde{s}}_j )$
does not contain the cavity mode. Next, we introduce two new collective degrees of freedom that connect the oriented and
the anti-oriented molecular subsystems, namely, a symmetric linear combination 
$\hat{r}=(\hat{X}+\hat{S})/\sqrt{2}$ and an anti-symmetric one 
$\hat{R}=(\hat{X}-\hat{S})/\sqrt{2}$. The symmetric coordinate $\hat{r}$ decouples, 
and the collective coupling of the whole isotropic ensemble to the cavity is captured by
the anti-symmetric coordinate $\hat{R}$. The coupling of $\hat{R}$ to the cavity has exactly
the same form as the Hamiltonian $\hat{\mathcal{H}}_{\rm col}$ [see Eq.~\eqref{eq:Hcmframe}] in the fully oriented ensemble,
\begin{equation}
    \hat{\mathcal{H}}_{\rm col\text{-}iso} = 
    - \frac{\partial^2_R}{2m}+\frac{m\Omega_\rmv^2}{2}\hat{R}^2 
    - \frac{\omega \partial^2_q}{2}
    + \frac{\omega}{2} \left(\hat{q} - g\sqrt{2N}\hat{R}\right)^2.
\end{equation}
From the above result, it becomes evident that the cavity-induced dynamics that we uncovered
for the case of the fully aligned ensemble, all transfer straightforwardly to the isotropic
ensemble as well, for the anti-symmetrically correlated coordinate between the anti-aligned
subsystems. Thus, the cavity-induced resonant and collective effects do not vanish for an
isotropic ensemble coupled to the cavity. The explicit dynamical behavior of the oriented
and the anti-oriented subsystems under different initial conditions will be discussed in a
future publication.

\textit{Discusion and Outlook.} A fundamental mechanistic understanding
of the observed VSC effects in polaritonic
chemistry would be a significant leap forward, as it would provide a minimal-waste form of
heterogeneous catalysis~\cite{JoelReview}. 

We propose a unique and general mechanism for modifying local molecular vibrations under
resonant collective VSC in a cavity, particularly in the limit of many molecules.
We discover a cavity-induced collective and resonant beating phenomenon that emerges in
the strong coupling regime, heralding significant modifications in the local vibrational
dynamics of molecules; excited in the intermediate out-of-equilibrium complexes formed
during chemical reactions. 

In our model, we assume that during a chemical reaction, a small fraction of reactant molecules forms an activated complex in which a specific bond is initially displaced from its equilibrium position. Under these initial conditions the collective motion of the system exhibits beatings that periodically suppress the motion of individual bonds. Consequently, vibrational energy is suppressed, limiting the complex’s ability to rearrange along other reaction pathways. If the beating period is comparable to the lifetime of the activated complex, the likelihood of it converting into products decreases, resulting in a measurable slowdown in the reaction rate. This effect spreads through the entire ensemble via cavity-mediated energy exchange, even when only a small fraction ($1-5\%$) of molecules are initially excited.

These phenomena are driven by strong cavity-molecule coupling,
resulting in polariton formation, which highlights the intricate interplay between excited
molecular ensembles and the cavity environment. The beating period is inversely proportional
to the VRS and thus depends on the number of collectively coupled molecules. The beating period exhibits 
a pronounced  peak at the cavity-molecule resonance. This is the origin of the
resonant suppression of local vibrations.

In the ultrastrong coupling regime, where the counter-rotating terms are important
~\cite{kockum2019ultrastrong, UltrastrongReview2019, QINultrastrong}, and cavity modifies the local
vibrations at short time scales, our theory makes an experimentally testable prediction:
under ultrastrong coupling, molecular vibrations can be modified off-resonance as
the beating period broadens around the cavity-molecule
resonance (see Fig.~\ref{fig:collective_vib}). This phenomenon could inspire new
experimental approaches and deepen our understanding of vibrational dynamics in cavities.
In addition, pulsed laser-assisted excitations in molecules can engineer
activated complexes~\cite{Nesbitt2000, NesbittReview}.  This straightforward method
offers a controlled way to test the proposed mechanism, potentially with cold and clean
molecular systems~\cite{DeMille2004MW, YeReview, tweezermolecules}.

A key aspect of our work is that the resonant suppression of vibrations requires
{\it only} a small fraction of excited molecules in the ensemble ($\sim 1\text{\,-\,}5\%$).
Importantly, the energy flows from the initially excited molecules into the ground-state
molecules, which then start to oscillate. Thus, the cavity mediates an energy exchange
and interactions between the excited and ground-state molecules. 

A comprehensive understanding of the cavity-induced modification of chemical reactivity
and dissociation dynamics requires the anharmonicity of the molecular potential to be included.
We intend to investigate the dynamics of Morse oscillators~\cite{Morse}
collectively coupled to the cavity mode in an upcoming publication.
We anticipate that the resonant cavity-mediated energy transfer between the molecules that
we uncover here will substantially modify the dissociation dynamics in the collective regime.
It is important to highlight that the anharmonic terms will couple the collective
Hamiltonian $\hat{\mathcal{H}}_{\rm col}$ and inter-molecular bond lengths 
$\hat{\mathcal{H}}_{\rm rel}$, introducing an additional layer of complexity. Further, the inclusion of anharmonicities will introduce non-linear dynamical
phenomena which go beyond the present formalism. At the same time, for anharmonic systems,
the dynamics of first moments cannot be fully captured with the semi-classical approach,
because of the quantum revivals emerging on the long time scale, as mentioned
in Refs.~\cite{LindoyQuantum2023, HuoQuantum2023}. For anharmonic systems, only the early-time
dynamics can be approximately captured with the semi-classical approach~\cite{TannorBook}. 

At this point, we would like to mention a relevant work which also studied
VSC from an out-of-equilibrium perspective utilizing classical dynamics
~\cite{WangFlickYelin2022}. This numerical study considered a model including vibrational
anharmonicities and bending modes. It was found that the dissociation dynamics of $N$
perfectly oriented classical molecules can be suppressed by strong coupling to a cavity
~\cite{WangFlickYelin2022}. In connection with our results, the resonant cavity-induced
collective beatings were not reported there.  

The role of molecular orientation and disorder~\cite{SidlerDisorder, RibeiroDisorder}
within our model needs to be investigated further. We expect that the
cavity-induced beatings will emerge even in the presence of weak disorder because the light-matter
interaction will dominate. At the same time, from the generalization of
our model to isotropic ensembles we anticipate that for ensembles where
half of the molecules are oriented parallel and half anti-parallel to the cavity field
polarization will also experience collective beatings and under different
out-of-equilibrium initial conditions, VSC will modify the local molecular vibrations. In addition, we comment on the role of thermal disorder, which typically increases the random motion and energy distribution among
molecules. In a more random
state, we anticipate that the beatings will be suppressed, due to the suppression of the
collective vibrations. However, a conclusive answer requires further investigation,
as molecular orientations also depend on the thermal fluctuations
in the ensemble.

This study advances the understanding of VSC and its implications in polaritonic
chemistry. We underline the importance of out-of-equilibrium molecular states in
hybrid cavity-molecule systems for local vibrational dynamics in activated complexes.
Our findings suggest that the molecular out-of-equilibrium states seed cavity-induced
modulation of chemical reactivity under VSC, potentially explaining chemical behavior
in polaritonic systems. Further, we showcase that resonant dynamical phenomena can
be captured with classical formalism, thus opening a path for understanding resonant
collective phenomena in polaritonic chemistry with classical dynamics simulations.
We thus provide a foundation for further exploration of cavity-modulated chemical
reactions. The ability to experimentally verify these predictions opens avenues
for developing tailored chemical systems that leverage VSC to control chemical
reactions and properties.

\textbf{Acknowledgements.} I.T. and H.R.S. acknowledge support from the NSF through
a grant for ITAMP at Harvard University. I.T. was also supported by an NSF subcontract
No. 3357 at the Smithsonian Astrophysical Observatory. This research was supported
in part by grant NSF PHY-2309135 to the Kavli Institute for Theoretical Physics (KITP).


\bibliography{Bibliography}

\end{document}